\documentclass[12pt,a4paper]{article}
\usepackage[top=25truemm,bottom=35truemm,left=25truemm,right=25truemm]{geometry}
\usepackage{graphicx}
\usepackage{amsmath,amssymb}
\usepackage{multirow}
\usepackage{hhline}
\usepackage[small]{caption}

\begin{document}
\setlength{\baselineskip}{18pt}
\begin{titlepage}
\begin{flushright}
\begin{tabular}{l}
 SU-HET-02-2014\\
 IPMU14-0034\\
 EPHOU-14-002
\end{tabular} 
\end{flushright}

\vspace*{1.2cm}
\begin{center}
{\Large\bf Accurate renormalization group analyses in neutrino sector}
\end{center}
\lineskip .75em
\vskip 1.5cm

\begin{center}
{\large Naoyuki Haba}$^1$,
{\large Kunio Kaneta}$^2$,
{\large Ryo Takahashi}$^1$, and
{\large Yuya Yamaguchi}$^3$\\

\vspace{1cm}

$^1${\it Graduate School of Science and Engineering, Shimane University,\\
 Matsue 690-8504, Japan}\\
 $^2${\it Kavli IPMU (WPI), The University of Tokyo,\\
 Kashiwa, Chiba 277-8568, Japan}\\
$^3${\it Department of Physics, Faculty of Science, Hokkaido University,\\
 Sapporo 060-0810, Japan}\\

\vspace{15mm}
{\bf Abstract}\\[5mm]
{\parbox{13cm}{\hspace{5mm}
%%%%%%%%%%%%%%%%%%%%%%%%%%%%%%%%%%%%%%%%%%%%%%%%%%%%%%%%%%%%%%%%%%%%%%%%%%%
%             ABSTRACT                      %
%%%%%%%%%%%%%%%%%%%%%%%%%%%%%%%%%%%%%%%%%%%%%%%%%%%%%%%%%%%%%%%%%%%%%%%%%%%
We investigate accurate renormalization group analyses in neutrino sector
 between $\nu$-oscillation and seesaw energy scales.
We consider decoupling effects of top quark and Higgs boson
 on the renormalization group equations of light neutrino mass matrix.
Since the decoupling effects are given in the standard model scale
 and independent of high energy physics,
 our method can basically apply to any models beyond the standard model.
We find that the decoupling effects of Higgs boson are negligible,
 while those of top quark are not.
Particularly, the decoupling effects of top quark affect neutrino mass eigenvalues,
 which are important for analyzing predictions such as mass squared differences
 and neutrinoless double beta decay in an underlying theory existing at high energy scale.
}}
\end{center}
\end{titlepage}

%%%%%%%%%%%%%%%%%%%%%%%%%%%%%%%%%%%%%%%%%%%%%%%%%%%%%%%%%%%%%%%%%%%%%%%%%%%
\section{Introduction}
%%%%%%%%%%%%%%%%%%%%%%%%%%%%%%%%%%%%%%%%%%%%%%%%%%%%%%%%%%%%%%%%%%%%%%%%%%%
Neutrino oscillation experiments established that active neutrinos are massive,
 and the masses are much smaller than the other standard model (SM) fermions.
The existence of nonzero neutrino masses is evidence of physics beyond the SM.
It is therefore necessary to explain the nonzero and tiny neutrino masses.
The seesaw mechanism\cite{seesaw} provides an attractive explanation,
 and a number of works have been presented in the context of the mechanism.
Moreover, recent precision measurements of leptonic mixing angles
 in the Pontecorvo-Maki-Nakagawa-Sakata (PMNS) matrix~\cite{Maki:1962mu}
 showed that $\theta_{12}$ and $\theta_{23}$ are large,
 and $\theta_{13}$ is small but not zero~\cite{An:2012eh,tortola,GonzalezGarcia:2012sz}.
These results suggest the mixing angles are much larger than mixing angles of quark sector.
Therefore, the nature of the neutrino is a key to study physics beyond the SM.

We can obtain some physical values in arbitrary high energy scale
 by solving the renormalization group equations (RGEs)
 and taking the experimental values as boundary conditions.
The renormalization group (RG) evolution of the light neutrino mass matrix
 can be determined by solving the RGE of a coefficient of effective dimension five operator
%  which is called Weinberg operator
 \cite{Weinberg:1979sa}.
The RG analyses using the operator are relevant only below the lowest seesaw scale,
 e.g., the lightest right-handed neutrino mass in type-I seesaw mechanism.
However, since the analyses are independent of the models,
 the analyses are useful for building models in high energy scale,
 in which the models are the grand unified theory (GUT),
 and/or have a new symmetry such as a flavor symmetry.
In fact, a large number of works respect with the RGEs of the neutrino sector have been presented
 (e.g., see \cite{Chankowski:1993tx}-\cite{degenerate}).
In particular, the RG effects can be large if the neutrino masses are quasi-degenerate
 \cite{Ellis:1999my,Haba:2000tx,Haba:2012ar,degenerate}.
There are also RG analyses in the minimal supersymmetric standard model (MSSM),
 which can realize the gauge coupling unification and be related to the GUT.

On the other hand, most of the analyses do not consider the decoupling effects of the massive SM particles.
When a certain particle is decoupled, contributions from the particle should be subtracted from the RGEs.
However, the decoupling effects are independent of the models beyond the SM,
 since the decoupling effects are of course given in the SM scale.
Thus, when we analyze the RG evolution in the MSSM,
 we should use the subtracted RGEs in the SM scale,
 while can use the original RGEs in the MSSM scale.
This method can basically apply to the other models beyond the SM.
In this paper, we consider the RGEs in the SM and the MSSM,
 and investigate the decoupling effects of top quark and Higgs boson
 on the light neutrino mass matrix
 between $\nu$-oscillation and seesaw energy scales.
The relevant RGEs of the work will be shown in Appendix.

In our analyses, the light neutrino mass matrix is approximately described only by two parameters.
One is an overall factor of the mass matrix,
 and the other denotes the RG effects of charged lepton Yukawa couplings
 and affects on the mixing angles.
We will show the RG evolution of these parameters in both the SM and MSSM,
 and find the decoupling effects are negligible for the latter parameter,
 while not for the former parameter.
Moreover, we will find that the effects are almost completely given by top quark decoupling,
 and the decoupling effects of Higgs boson are negligible.
In the MSSM, these fundamental behaviors are the same as in the SM.
Besides, when $\tan \beta \simeq 1$, the RG evolution is similar to the SM results.
Next, we will show the RG evolution of the mass squared differences and the mixing angles,
 in which the results are correspond to the MSSM with $\tan \beta = 30$.
% Since the RG effects of the mixing angles in the SM or the MSSM with small $\tan \beta$ are small,
%  we do not show these results.
We will find the decoupling effects are negligible for the mixing angles,
 while not for the mass eigenvalues.
These results are important for analyzing predictions such as mass squared differences
 and neutrinoless double beta decay in an underlying theory existing at high energy scale.
We will also discuss the dependence of decoupling effects on mass spectrum of light neutrinos, degeneracy of the masses and CP-phases.
% These results will be shown in Sec.\,\ref{sec3},
%  and the relevant RGEs of the work will be shown in Appendix.

%%%%%%%%%%%%%%%%%%%%%%%%%%%%%%%%%%%%%%%%%%%%%%%%%%%%%%%%%%%%%%%%%%%%%%%%%%%
\section{Renormalization Group Evolution of Neutrino\\ Mass Matrix}
%%%%%%%%%%%%%%%%%%%%%%%%%%%%%%%%%%%%%%%%%%%%%%%%%%%%%%%%%%%%%%%%%%%%%%%%%%%

%%%%%%%%%%%%%%%%%%%%%%%%%%%%%%%%%%%%%%%%%%%%%
\subsection{Neutrino Mass Matrix} \label{sec2}
%%%%%%%%%%%%%%%%%%%%%%%%%%%%%%%%%%%%%%%%%%%%%
We consider the extensions of the SM and the MSSM, in which lepton mass terms
 in low energy scale are effectively given by
\begin{eqnarray}
	&&\mathcal{L_\nu} = - Y_E \bar{L} \Phi E_R - \frac{\kappa}{2} (\overline{L^C} \Phi) (L \Phi) + {\rm h.c.}\,, %\quad {\rm for\ the\ SM},\\
% 	&&\mathcal{L_\nu} = - Y_E \bar{L} \Phi E_R - \frac{\kappa}{2} (L \Phi_u) (L \Phi_u) + {\rm h.c.} \quad {\rm for\ the\ MSSM},
\end{eqnarray}
where $Y_E$, $L$, $E_R$, and $\Phi$ are the Yukawa coupling matrix of charged leptons,
 left-handed lepton doublets, right-handed charged leptons, and (up-type) Higgs doublet
 in the SM (the MSSM), respectively. 
$\kappa$ is a coefficient of effective dimension five operator. %which is called Weinberg operator.
Now an effective light neutrino mass matrix $M_\nu$ is given by $\kappa v^2$,
 where $v$ is a relevant Higgs vacuum expectation value,
 that is, $v = 174$ GeV in the SM and $v = 174 \times \sin \beta$ GeV in the MSSM, respectively.

On the other hand, the light neutrino mass matrix can also be described
 by the PMNS matrix $U$ and mass eigenvalues of light neutrinos:
\begin{eqnarray} \label{lM}
	(M_\nu)_{\alpha \beta} = (U^* M_\nu^{\rm diag} U^\dagger)_{\alpha \beta}
									= (U^* \cdot {\rm Diag} \{m_1,m_2,m_3\} \cdot U^\dagger)_{\alpha \beta}
									= \sum_i U^*_{\alpha i} U^*_{\beta i} m_i \ ,
\end{eqnarray}
where the charged lepton mass matrix is diagonal,
 and $M_\nu^{\rm diag}$ is a diagonal matrix, and $\alpha, \beta=e, \mu, \tau$.
Then, if the neutrinos are Majorana particles,
 the mass matrix can be described by 3 mixing angles, 3 mass eigenvalues of the neutrinos
 and 3 CP-phases (one Dirac phase and two Majorana phases),
  in which $U$ is written by
\begin{eqnarray}
	U = \left( \begin{array}{ccc}
		c_{12} c_{13} & s_{12} c_{13} & s_{13} e^{- i \delta} \\
		-s_{12} c_{23} - c_{12} s_{23} s_{13} e^{i \delta} & c_{12} c_{23} - s_{12} s_{23} s_{13} e^{i \delta} & s_{23} c_{13} \\
		s_{12} s_{23} - c_{12} c_{23} s_{13} e^{i \delta} & - c_{12} s_{23} - s_{12} c_{23} s_{13} e^{i \delta} & c_{23} c_{13}
		\end{array} \right)
		\left( \begin{array}{ccc}
		e^{- i \frac{\phi_1}{2}} & 0 & 0 \\
		0 & e^{- i \frac{\phi_2}{2}}& 0 \\
		0 & 0 & 1
		\end{array} \right) .
\end{eqnarray}
Once one fixes those values at low energy as boundary conditions,
 one can obtain those values at arbitrary high energy scale
 by solving the corresponding RGEs.

The RGE for $\kappa \equiv M_\nu / v^2$ is given by
\begin{eqnarray}
	16 \pi^2 \frac{{\rm d} \kappa}{{\rm d} t} = C_E (Y_E^\dagger Y_E)^T \, \kappa + C_E \,\kappa \, (Y_E^\dagger Y_E)
		 + \bar{\alpha} \, \kappa \,,
\end{eqnarray}
with $t \equiv  \ln \mu$ ($\mu$ is a renormalization scale), where $C_E = -3/2$ in the SM and $C_E = 1$ in the MSSM, respectively.
And,
\begin{eqnarray}
	\bar{\alpha}_{{\rm SM}} &=& 2 \, {\rm Tr} \left[3 Y_U^\dagger Y_U + 3 Y_D^\dagger Y_D + Y_E^\dagger Y_E \right] - 3 g_2^2 + \lambda \, ,\\
	\bar{\alpha}_{{\rm MSSM}} &=& 6 \, {\rm Tr} \left[Y_U^\dagger Y_U \right] - \frac{6}{5} g_1^2 - 6 g_2^2 \, , \label{al}
\end{eqnarray}
at one-loop level,
 where $Y_f \, (f \in \{ E,U,D \})$ are Yukawa coupling matrices of the charged leptons, up- and down-type quarks, respectively,
 $g_i$ are gauge coupling constants and $\lambda$ is the Higgs self coupling in the SM.
Then, we can write the neutrino mass matrix as
 $M_\nu (\Lambda) = R \, ( \, I \, M_\nu (\Lambda_{\rm EW}) \, I \, )$ at arbitrary high energy scale $\Lambda$,
 where $\Lambda_{{\rm EW}}$ is some energy at electroweak scale, $R$ is a flavor blind overall factor,
 and $I$ is defined by $I^{-1} \equiv {\rm Diag} \{ \sqrt{I_e}, \sqrt{I_\mu}, \sqrt{I_\tau} \}$~\cite{Ellis:1999my}-\cite{Haba:1999fk}. 
$I_\alpha$ denote quantum corrections of the charged lepton Yukawa couplings
 as $I_\alpha \equiv \exp \left[ - \frac{C_E}{8 \pi^2} \int_{t_{\rm EW}}^{t_\Lambda} dt \, y_\alpha^2 \right]$
 with $t_\Lambda \equiv \ln \Lambda$ and $t_{\rm EW} \equiv \ln \Lambda_{\rm EW}$.
Then, the light neutrino mass matrix at arbitrary high energy scale can be written by
\begin{eqnarray}
	M_\nu (\Lambda) = r \left( \begin{array}{ccc}
		(M_\nu(\Lambda_{\rm EW}))_{ee} & (M_\nu(\Lambda_{\rm EW}))_{e\mu} \, \sqrt{\frac{I_e}{I_\mu}} & (M_\nu(\Lambda_{\rm EW}))_{e\tau} \, \sqrt{\frac{I_e}{I_\tau}} \\
		(M_\nu(\Lambda_{\rm EW}))_{e\mu} \, \sqrt{\frac{I_e}{I_\mu}} & (M_\nu(\Lambda_{\rm EW}))_{\mu\mu} \, \frac{I_e}{I_\mu} & (M_\nu(\Lambda_{\rm EW}))_{\mu\tau} \, \sqrt{\frac{I_e}{I_\mu} \, \frac{I_e}{I_\tau}} \\
		(M_\nu(\Lambda_{\rm EW}))_{e\tau} \, \sqrt{\frac{I_e}{I_\tau}} & (M_\nu(\Lambda_{\rm EW}))_{\mu\tau} \, \sqrt{\frac{I_e}{I_\mu} \, \frac{I_e}{I_\tau}} & (M_\nu(\Lambda_{\rm EW}))_{\tau\tau} \, \frac{I_e}{I_\tau}
	\end{array} \right), \label{hM}
\end{eqnarray}
 where $r \equiv R/I_e$.
Now we introduce small parameters defined as $\epsilon_\tau \equiv \sqrt{I_e / I_\tau} - 1$ and 
$\epsilon_\mu \equiv \sqrt{I_e / I_\mu} - 1$.
Since $\epsilon_\mu \ll \epsilon_\tau$ and $\epsilon_\mu$ is numerically almost equal to 0,
 we can neglect $\epsilon_\mu$.
Thus, Eq.\,(\ref{hM}) can be well approximated by
\begin{eqnarray}
	M_\nu (\Lambda) \simeq r \left( \begin{array}{ccc}
     (M_\nu(\Lambda_{\rm EW}))_{ee} & (M_\nu(\Lambda_{\rm EW}))_{e\mu} & (M_\nu(\Lambda_{\rm EW}))_{e\tau} \, (1+\epsilon) \\
     (M_\nu(\Lambda_{\rm EW}))_{e\mu} & (M_\nu(\Lambda_{\rm EW}))_{\mu\mu} & (M_\nu(\Lambda_{\rm EW}))_{\mu\tau} \, (1+\epsilon) \\
     (M_\nu(\Lambda_{\rm EW}))_{e\tau} \, (1+\epsilon) & (M_\nu(\Lambda_{\rm EW}))_{\mu\tau} \, (1+\epsilon) & (M_\nu(\Lambda_{\rm EW}))_{\tau\tau} \, (1+\epsilon)^2
	\end{array} \right), \label{hMa}
\end{eqnarray}
 where we drop the subscript of $\epsilon_\tau$, that is, $\epsilon \equiv \epsilon_\tau$.
To investigate the RG evolution of the mass matrix,
 all we have to do is calculating $r$ and $\epsilon$ at arbitrary energy scale.
$r$ is calculated by
\begin{eqnarray}
	r (\Lambda) = \frac{(M_\nu(\Lambda))_{ee}}{(M_\nu(\Lambda_{{\rm EW}}))_{ee}} \, ,\label{r}
\end{eqnarray}
and $\epsilon$ is calculated by
\begin{eqnarray}
	\epsilon (\Lambda) = \sqrt{ \frac{I_e}{I_\tau}} - 1
								= \exp \left[ \frac{1}{2} \frac{C_E}{8 \pi^2} \int_{t_{\rm EW}}^{t_\Lambda} dt \, (y_\tau^2 - y_e^2) \right] -1 \, .\label{epsilon}
\end{eqnarray}
The mass eigenvalues and the mixing angles can be extracted from the mass matrix.
Note that the mass eigenvalues depend on both $r$ and $\epsilon$,
 while the mixing angles depend only on $\epsilon$.

%%%%%%%%%%%%%%%%%%%%%%%%%%%%%%%%%%%%%%%%%%%%%
\subsection{Treatment of Decoupling Effects} \label{decoupling}
%%%%%%%%%%%%%%%%%%%%%%%%%%%%%%%%%%%%%%%%%%%%%
In addition to the above discussion,
 we should consider decoupling effects of the massive SM particles at low energy scale.
Among the SM particles the order of their masses is $m_t^{\rm pole} > m_h > M_Z > \cdots$,
 where $m_t^{\rm pole}$, $m_h$, and $M_Z$ are pole mass of top quark, masses of Higgs boson and Z boson, respectively.
Thus, for $m_h \leq \mu < m_t^{\rm pole}$ top quark is decoupled,
 for $M_Z \leq \mu < m_h$ top quark and Higgs boson are decoupled, and so on.
When we solve the RGEs, in most cases we take the boundary conditions at $\mu = M_Z$.
Thus, we should consider the decoupling effects only of top quark and Higgs boson.
However, the decoupling effects are independent of the models beyond the SM,
 since the decoupling effects are of course given in the SM scale.
Therefore, when we analyze the RG evolution in the MSSM,
 we should use the subtracted RGEs in the SM scale,
 while can use the original RGEs in the MSSM scale.
This method can basically apply to the other models beyond the SM.
The relevant RGEs of the work are shown in Appendix.

\begin{figure}[t]
  \begin{center}
      \begin{minipage}{0.7\hsize}
        \begin{center}
          \includegraphics[clip, width=\hsize]{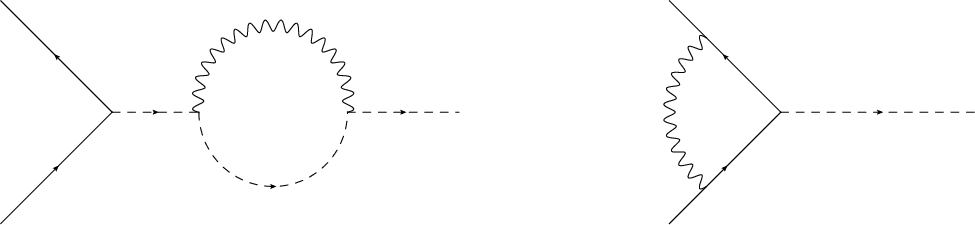}
        \end{center}
      \end{minipage}
    \caption{Diagrams which contribute to $\beta$-function of fermion Yukawa couplings.
				The solid, dashed, and wavy lines show fermions, Higgs boson, and gauge bosons, respectively.}
    \label{diagram}
  \end{center}
\end{figure}

Let us explain our treatment of the decoupling effects.
First, for $m_h \leq \mu < m_t^{\rm pole}$
 top quark is decoupled and does not appear as the internal line in Feynman diagrams.
So, we subtract the contributions of the corresponding diagrams of top quark loop.
The decoupling effects are shown as $-3 y_t^2$ or $-3 y_t^4$,
 which cancel top quark Yukawa coupling in Tr[$Y_U^\dagger Y_U$] or Tr[$Y_U^\dagger Y_U Y_U^\dagger Y_U$]
 in Eqs.(\ref{kappa_mt})-(\ref{lambda_mt}).
Therefore, $\beta$-functions do not include top quark Yukawa coupling for $\mu < m_t^{\rm pole}$.
Similarly, for $M_Z \leq \mu < m_h$
 Higgs boson also does not appear as the internal line in Feynman diagrams.
Then, $\beta$-function of $\kappa$ has only one term which is proportional to SU(2) gauge coupling,
 and $\beta$-function of $\lambda$ has only contributions of fermion box diagrams,
 which appear as fourth power of Yukawa couplings.
For $\beta$-functions of fermion Yukawa couplings,
 the terms of gauge couplings remain.
In order to calculate contributions of electroweak gauge bosons,
 we use Landau gauge, in which only two diagrams shown in Fig.\ref{diagram} have nonzero contributions.
Particularly, for $M_Z \leq \mu < m_h$
 we have to calculate only the right figure,
 which has U(1) gauge boson.
As a result, we obtain the RGEs given by Eqs.(\ref{kappa_mh})-(\ref{lambda_mh}).

Finally, we comment on matching conditions for the running couplings.
For example, $\kappa$ is sensitive to the decouplings of top, Higgs and SUSY particles at corresponding threshold scales,
 and thus the matchings at those thresholds should be considered.
So far, our analysis is up to 1-loop level, and we do not include threshold corrections on $\kappa$
 since they are typically smaller than 2-loop effects.
Therefore, we take a simple way in which the $\kappa$ running by the $\beta$-functions (3a), (4a) and (5a)
 is continuously connected at the thresholds without the corrections.
The treatment is the same for the other couplings except for top quark Yukawa coupling.
Since top quark Yukawa coupling is determined at the scale of the top pole mass,
 and thus we have set the matching condition including the threshold correction,
 which is given by $m_t^{\rm pole}=m_t(\mu=m_t^{\rm pole}) (1+\delta_{\rm th})$,
where  $m_t(\mu)$ and $\delta_{\rm th}$ denote the running top mass and whole 1-loop threshold corrections, respectively.
 
% \newpage
%%%%%%%%%%%%%%%%%%%%%%%%%%%%%%%%%%%%%%%%%%%%%%%%%%%%%%%%%%%%%%%%%%%%%%%%%%%
\section{Numerical Analyses of Neutrino Mass Matrix} \label{sec3}
%%%%%%%%%%%%%%%%%%%%%%%%%%%%%%%%%%%%%%%%%%%%%%%%%%%%%%%%%%%%%%%%%%%%%%%%%%%

%%%%%%%%%%%%%%%%%%%%%%%%%%%%%%%%%%%%%%%%%%%%%
\subsection{Boundary Conditions}
%%%%%%%%%%%%%%%%%%%%%%%%%%%%%%%%%%%%%%%%%%%%%
To solve the RGEs, we take the boundary conditions for fermions and bosons as
\begin{eqnarray*}
	m_u &=& 2.3 \, {\rm MeV}, \quad \quad m_c = 1.28 \, {\rm GeV},\\
	m_d &=& 4.8 \, {\rm MeV}, \quad \quad m_s = 95 \, {\rm MeV}, \quad \quad \ \ \ m_b = 4.18 \, {\rm GeV},\\
 	m_e &=& 0.511 \, {\rm MeV}, \quad m_\mu = 106 \, {\rm MeV}, \quad \quad \ m_\tau = 1.78 \, {\rm GeV},\\
	M_Z &=& 91.2 \, {\rm GeV}, \quad \ \  m_h = 126 \, {\rm GeV},\\
	\alpha_{em}^{-1} &=& 127.944\,, \quad \quad \sin^2 \theta_w = 0.23116\,,\quad \, \alpha_s \equiv g_3^2/(4 \pi) = 0.1184\,,
\end{eqnarray*}
at $\mu = M_Z$, and $m_t = 160$\,GeV at $\mu = m_t^{{\rm pole}} = 173$\,GeV \cite{Beringer:1900zz,Zhou}.
$\alpha_{em}$, $\theta_w$, and $g_3$ are fine-structure constant, weak mixing angle, and strong coupling constant, respectively.
The experimental values for the mass eigenvalues and the mixing angles in low energy scale are given by the best-fit values \cite{Capozzi:2013csa}:
\begin{table}[h]
  \begin{center}
	\begin{tabular}{|c|c|c|c|c|c|} \hline
		& $m_2^2 - m_1^2$ & $|m_3^2 - \frac{m_1^2 + m_2^2}{2}|$ & $\sin^2 \theta _{12}$ & $\sin^2 \theta _{23}$ & $\sin^2 \theta _{13}$  \\ \hline
		Best-fit & \multirow{2}{26mm}{$7.54 \times 10^{-5}$eV$^2$} & $2.44 \times 10^{-3}$ eV$^2$\,(NH) & \multirow{2}{10mm}{0.308} & 0.425\,(NH) & 0.0234\,(NH) \\
		values & & $2.40 \times 10^{-3}$ eV$^2$\,(IH) & & 0.437\,(IH) & 0.0239\,(IH) \\ \hline
	\end{tabular}
% \caption{boundary conditions}
\end{center}
\end{table}\\
We use these values as the boundary conditions at $\mu = M_Z$.
In fact, the $\beta$-function of $\kappa$ is zero below $\mu = M_Z$.
Therefore, our analyses including the decoupling effects can accurately connect $\nu$-oscillation to seesaw energy scale.

%%%%%%%%%%%%%%%%%%%%%%%%%%%%%%%%%%%%%%%%%%%%%
\subsection{RG Evolution of $r$ and $\epsilon$} \label{MainResults}
%%%%%%%%%%%%%%%%%%%%%%%%%%%%%%%%%%%%%%%%%%%%%
We show the RG evolution of $r$ and $\epsilon$ in this subsection.
In our notation, $r$ and $\epsilon$ are calculated by Eqs.(\ref{r}) and (\ref{epsilon}).
In this subsection, we consider the mass spectrum of light neutrinos is the NH and $m_1 = 0$\,eV,
 and all figures show within $M_Z \leq \mu \leq 10^{14}$\,GeV.
Since, when we consider the type-I seesaw mechanism, the neutrino Yukawa couplings exceed 1 at higher energy scale than $\mu = 10^{14}$\,GeV,
 we consider the lower energy scale than $\mu = 10^{14}$\,GeV.
And, since we take the boundary conditions of the RGEs at $\mu = M_Z$ (except for $m_t$),
 $r = 1$ and $\epsilon = 0$ at $\mu = M_Z$.

% \newpage
\begin{figure}[t]
  \begin{center}
      \begin{minipage}{0.46\hsize}
        \begin{center}
          \includegraphics[clip, width=\hsize]{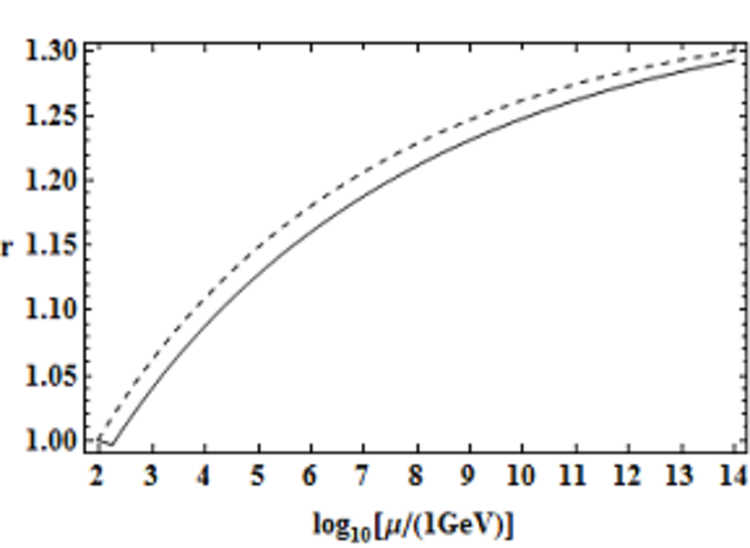}
        \end{center}
      \end{minipage}
\hspace{1mm}
      \begin{minipage}{0.51\hsize}
        \begin{center}
          \includegraphics[clip, width=\hsize]{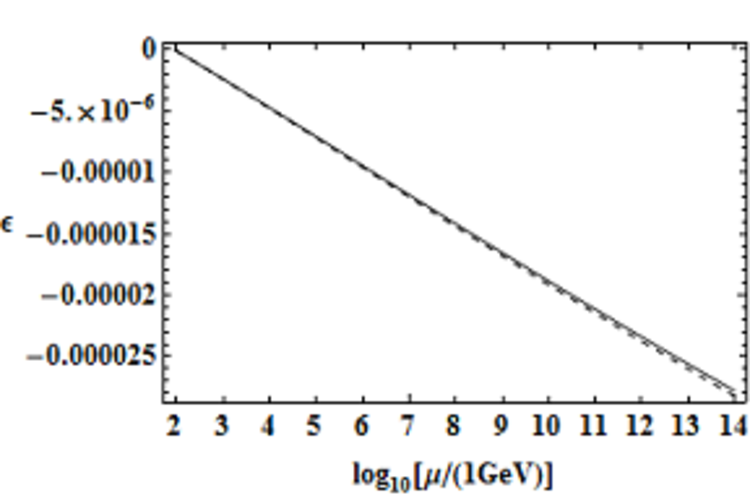}
        \end{center}
      \end{minipage}
  \end{center}
% \end{figure}
\vspace{-7mm}
% \begin{figure}[ht]
  \begin{center}
      \begin{minipage}{0.46\hsize}
        \begin{center}
          \includegraphics[clip, width=\hsize]{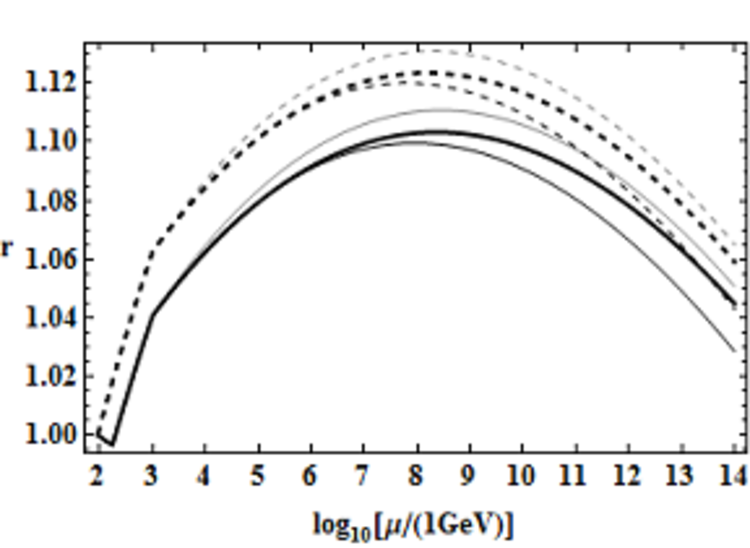}
        \end{center}
      \end{minipage}
\hspace{7mm}
      \begin{minipage}{0.46\hsize}
        \begin{center}
          \includegraphics[clip, width=\hsize]{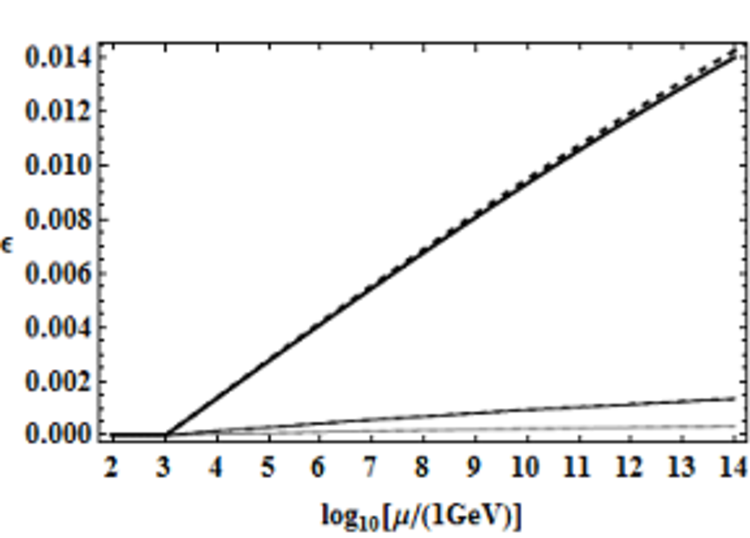}
        \end{center}
      \end{minipage}
    \caption{RG evolution of $r$ and $\epsilon$.
				The upper and lower figures show the results in the SM and the MSSM (SUSY threshold is taken at 1\,TeV), respectively.
				The solid and dashed lines show the results including the decoupling effects and not, respectively.
				The gray, black, and black-thick lines represent $\tan \beta = 5$, $\tan \beta = 10$, and $\tan \beta = 30$, respectively.}
    \label{r_epsilon}
  \end{center}
\end{figure}

The upper figures of Fig.\ref{r_epsilon} show the RG evolution of $r$ and $\epsilon$ in the SM.
We can see that the decoupling effects of top quark and Higgs boson
 are negligible for $\epsilon$, but for $r$.
For $r$, the difference between including the decoupling effects or not
 is specifically about 0.6$\%$ at $\mu = 10^{14}$\,GeV.
In fact, the decoupling effects of Higgs boson are negligible.
Thus, the top quark decoupling accounts for the difference,
 since top quark Yukawa coupling is much larger than the others.
The sign inversion of $r$ at $\log m_t^{\rm pole} \simeq 2.2$ just occur due to the top quark decoupling,
 that is, the sign of $\beta$-function of $\kappa$ is changed
 when the contributions from top quark are subtracted from the RGEs.
On the other hand, since $\epsilon$ depends on the integral of charged lepton Yukawa couplings,
 the decoupling effects are buried in the integrated value,
 that is, the decoupling effects are negligible.

The lower figures of Fig.\ref{r_epsilon} show the RG evolution of $r$ and $\epsilon$ in the MSSM.
The gradient of $r$ in high energy scale is positive in the SM, but negative in the MSSM,
 since top quark Yukawa coupling has positive contribution to the $\beta$-function and becomes dominant below $\mu = {\cal O}(10^{8 - 9})$\,GeV,
 while gauge couplings have negative contribution and become dominant above $\mu = {\cal O}(10^{8 - 9})$\,GeV.
The gradient of $\epsilon$ is negative in the SM, but positive in the MSSM due to the sign of $C_E$.
We can see that the values in the MSSM scale depend on $\tan \beta$.
But, the differences between including the decoupling effects or not
 are almost independent of $\tan \beta$.
The differences are about 1.4$\%$ for $r$ at $\mu = 10^{14}$\,GeV, and negligible for $\epsilon$ in all energy scale.

\begin{figure}[t]
  \begin{center}
      \begin{minipage}{0.45\hsize}
        \begin{center}
          \includegraphics[clip, width=\hsize]{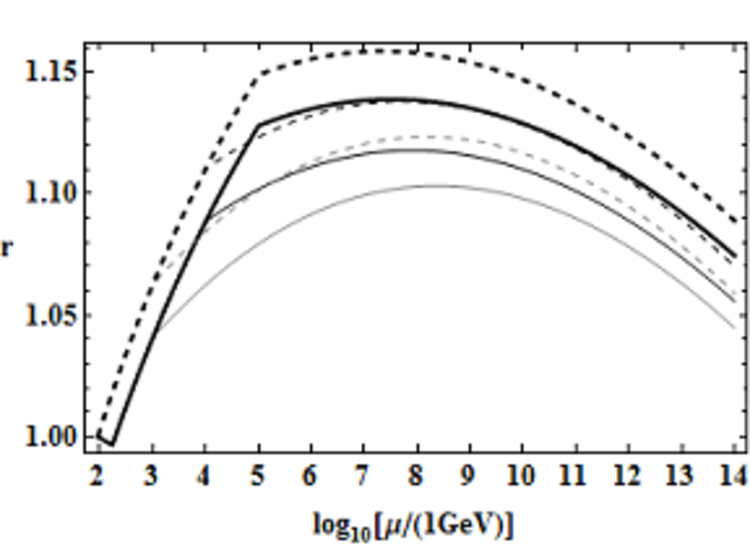}
        \end{center}
      \end{minipage}
\hspace{6mm}
      \begin{minipage}{0.46\hsize}
        \begin{center}
          \includegraphics[clip, width=\hsize]{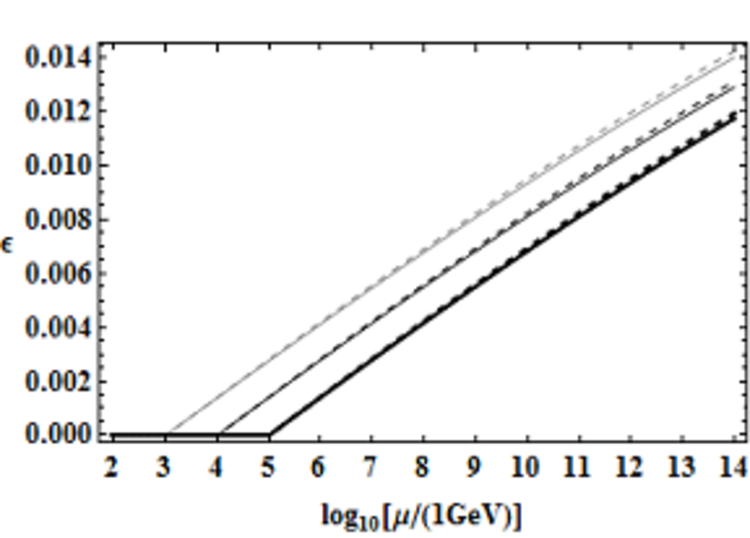}
        \end{center}
      \end{minipage}
    \caption{SUSY threshold dependence of $r$ and $\epsilon$ in the MSSM with $\tan \beta = 30$.
				The solid and dashed lines show the results including the decoupling effects and not, respectively.
				The gray, black, and black-thick lines represent the cases that
				SUSY threshold are taken at $\mu =$ 1\,TeV, 10\,TeV, and 100\,TeV, respectively.}
    \label{NH_tanb30}
  \end{center}
\end{figure}

\begin{figure}[t]
  \begin{center}
      \begin{minipage}{0.47\hsize}
        \begin{center}
          \includegraphics[clip, width=\hsize]{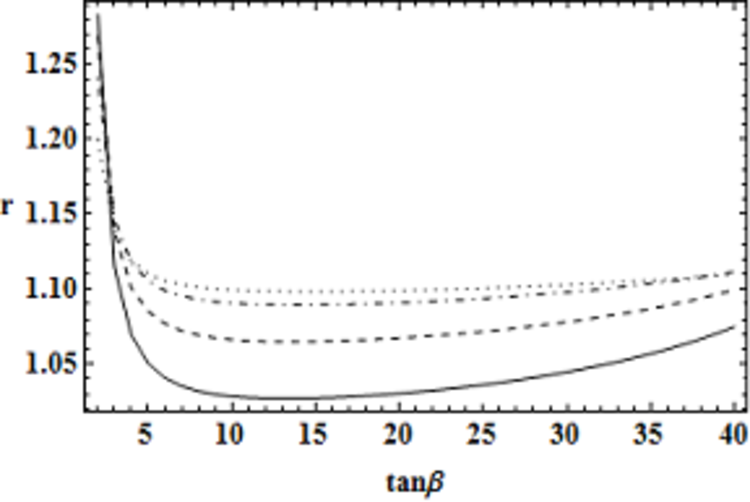}
        \end{center}
      \end{minipage}
\hspace{5mm}
      \begin{minipage}{0.48\hsize}
        \begin{center}
          \includegraphics[clip, width=\hsize]{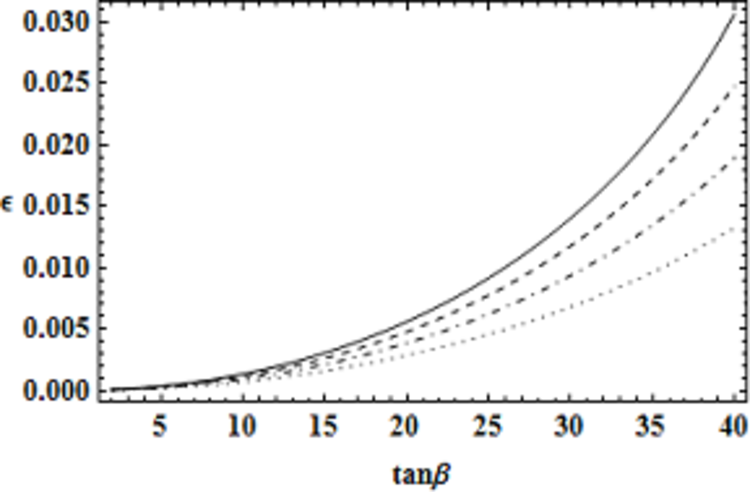}
        \end{center}
      \end{minipage}
    \caption{$\tan \beta$ dependence of $r$ and $\epsilon$ in the MSSM with the decoupling effects.
				SUSY threshold is taken at $\mu =$1\,TeV.
				The dotted, dot-dashed, dashed, and solid lines represent values
				at $\mu = 10^{8}$\,GeV, $10^{10}$\,GeV, $10^{12}$\,GeV, and $10^{14}$\,GeV, respectively.}
    \label{tanb}
  \end{center}
\end{figure}

Figure \ref{NH_tanb30} shows the SUSY threshold dependence of $r$ and $\epsilon$ in the MSSM with $\tan \beta = 30$.
The fundamental behaviors are the same as before.
We can see that the differences between including the decoupling effects or not
 are almost independent of the value of SUSY threshold.

Figure \ref{tanb} shows the $\tan \beta$ dependence of $r$ and $\epsilon$ in the MSSM with the decoupling effects.
When $\tan \beta \simeq 1$, the RG evolution is similar to the SM results.
We can see that the minimal RG effect of $r$ occurs at $\tan \beta \simeq 13$,
 and $\epsilon$ can be large for large $\tan \beta$.
The reason for $\epsilon$ is simply because charged lepton Yukawa couplings are larger as $\tan \beta$ is large.
Moreover, top quark Yukawa coupling is smaller, while bottom quark Yukawa coupling is larger.
Then, top quark Yukawa coupling accidentally has the minimum at $\tan \beta \simeq 13$.
This is the reason for $r$.
Note that since the RG effects of the mixing angles depend only on $\epsilon$, 
 the mixing angles can be unstable for large $\epsilon$ as we will show the next subsection.

In Figs.\ref{r_epsilon}\,-\,\ref{tanb}, we have considered the mass spectrum of light neutrinos as the NH and $m_1 = 0$\,eV.
Note that all figures are the same even if the mass spectrum is the IH,
 or the lightest neutrino mass is large as 0.05\,eV, that is, the masses are quasi-degenerate.
When we change the mass spectrum or the absolute neutrino mass,
 the light neutrino mass matrix (equivalently $\kappa$) also changes.
But, $r$, which is proportional to the ratio of $\kappa$, does not depend on the magnitude of $\kappa$,
 since the magnitude is canceled in the ratio.
%  the $\beta$-function of $\kappa$ is proportional to $\kappa$,
$\epsilon$ obviously does not depend on the magnitude of $\kappa$,
 since $\epsilon$ is calculated by charged lepton Yukawa couplings.
Moreover, both $r$ and $\epsilon$ are independent of CP-phases,
 because the arguments of $r$ and $\epsilon$ do not change during the RG evolution.

We note that the effective neutrino mass $(M_\nu)_{11}$ is given by $r \times (M_\nu (M_Z))_{11}$.
The amplitude of neutrinoless double beta decay is proportional to $(M_\nu)_{11}$.
Therefore, we can easily see the RG evolution of the decay amplitude.
On the other hand, experiments of the decay can restrict the absolute neutrino mass scale.
Since we often consider the neutrino mass scale relates to unknown high energy physics,
 the RG evolution is important for constructing the models in high energy scale.
Similarly, our analyses are useful for the other phenomenological problems,
 e.g., thermal leptogenesis\cite{fy86}, which is proposed to explain the baryon asymmetry in the universe.
 In the leptogenesis, the heaviest mass eigenvalue and the absolute neutrino mass are important parameters
 used to calculate the baryon asymmetry \cite{Davidson:2002qv}.
Since the mass eigenvalues are obtained by $r$ and $\epsilon$, and almost depend on $r$,
 the decoupling effects are not negligible.
Thus, our results for the neutrino mass might become important
 for accurate computation in the canonical leptogenesis.

% \newpage

\begin{figure}[ht]
  \begin{center}
      \begin{minipage}{0.4\hsize}
        \begin{center}
          \includegraphics[clip, width=\hsize]{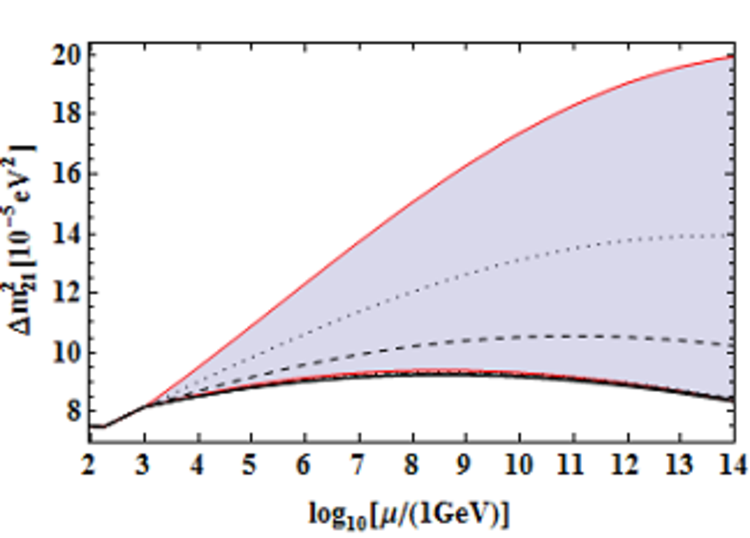}
        \end{center}
      \end{minipage}
\hspace{10mm}
      \begin{minipage}{0.4\hsize}
        \begin{center}
          \includegraphics[clip, width=\hsize]{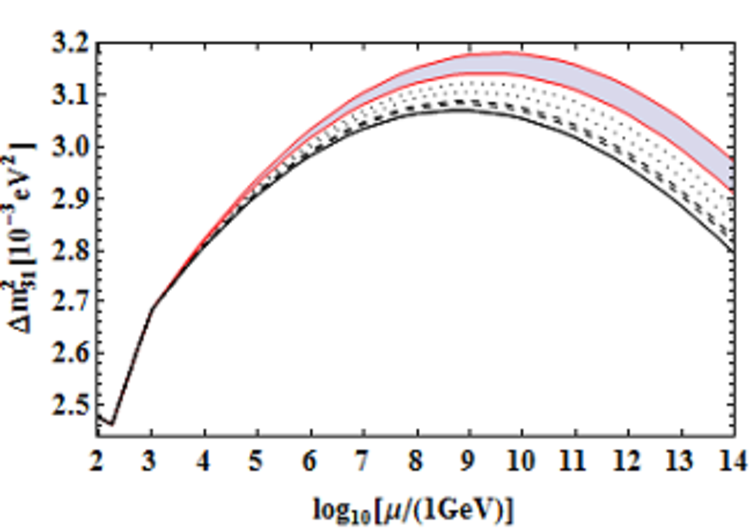}
        \end{center}
      \end{minipage}
  \end{center}
% \end{figure}
\vspace{-9mm}
% \begin{figure}[ht]
  \begin{center}
      \begin{minipage}{0.4\hsize}
        \begin{center}
          \includegraphics[clip, width=\hsize]{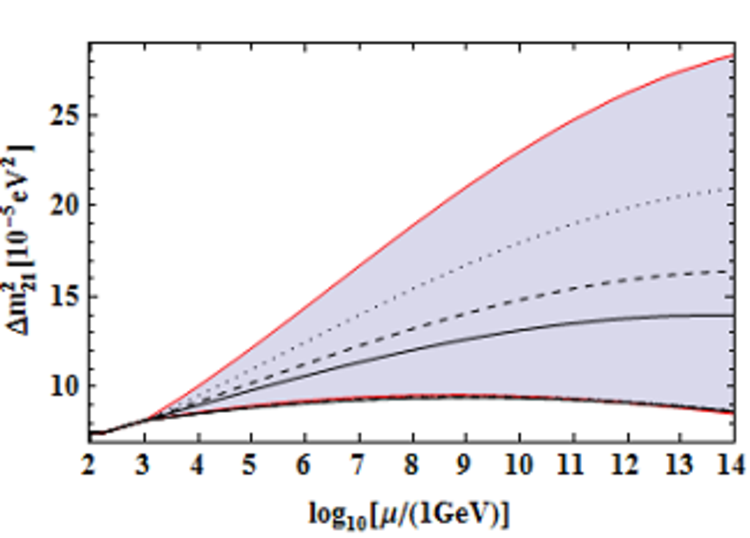}
        \end{center}
      \end{minipage}
\hspace{10mm}
      \begin{minipage}{0.4\hsize}
        \begin{center}
          \includegraphics[clip, width=\hsize]{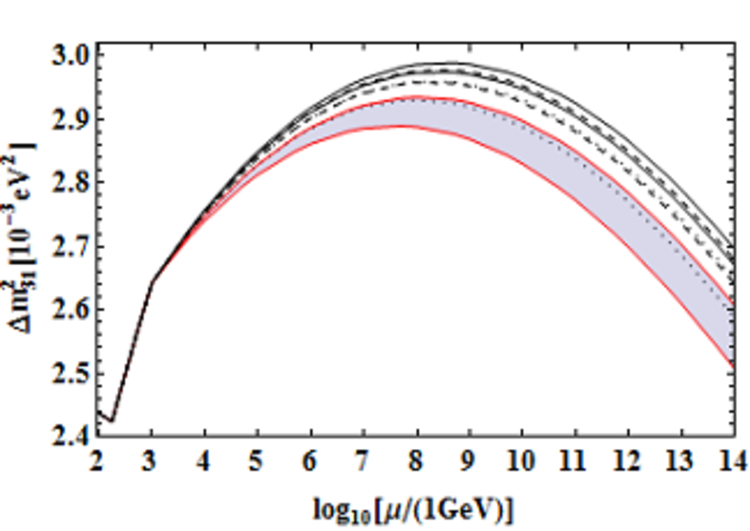}
        \end{center}
      \end{minipage}
    \caption{RG evolution of $\Delta m_{21}^2$ and $\Delta m_{31}^2$ in the MSSM with the decoupling effects.
				The upper (lower) figures show the results in the NH (IH).
				The solid, dashed, dotted, and red-solid lines correspond to the upper and lower bound of allowed region
				 for $m_{1 {\rm \,(or\,3)}}$ = 0\,eV, 0.03\,eV, 0.05\,eV, and 0.07\,eV, respectively.
				The shaded regions can be taken according to CP-phases for $m_{1 {\rm \,(or\,3)}}$ = 0.07\,eV.}
    \label{deltam}
  \end{center}
\end{figure}

\begin{table}[!ht]
  \begin{center}
  	\begin{tabular}{|c||c|c|} \hline
		NH & $\Delta m_{21}^2$ & $\Delta m_{31}^2$ \\ \hline \hline
		Upper bound & \qquad\, (0, any, $\pi$)\qquad / (0, $\pi$, $\pi$) & \qquad\, (0, any, 0)\qquad / (0, 0, 0) \\ \hline
		Lower bound & \qquad\, ($\pi$, any, 0)\qquad / ($\pi$, $\pi$, 0) & \qquad\, (0, any, $\pi$)\qquad / ($\pi$, $\pi$, 0) \\ \hline
	\end{tabular}\\
\vspace{2mm}
	\begin{tabular}{|c||c|c|} \hline
		IH & $\Delta m_{21}^2$ & $\Delta m_{31}^2$ \\ \hline \hline
		Upper bound & $\delta = 0, |\phi_1 - \phi_2| = 0$ / (0, 0, 0) & $\delta = \pi, |\phi_1 - \phi_2| = \pi$ / ($\pi$, 0, $\pi$) \\ \hline %\hline
		Lower bound & $\delta = \pi, |\phi_1 - \phi_2| = \pi$ / ($\pi$, 0, $\pi$) & $\delta = 0, |\phi_1 - \phi_2| = 0$ / (0, $\pi$, $\pi$) \\ \hline
	\end{tabular}
	\caption{Combinations of CP-phases which give the upper and lower bounds of $\Delta m_{21}^2$ and $\Delta m_{31}^2$.
	The values in the table are ($\delta, \phi_1, \phi_2$),
	 and the former and latter combinations correspond to $m_{1 {\rm \,(or\,3)}}$ = 0\,eV and nonzero $m_{1 {\rm \,(or\,3)}}$, respectively.
	The upper (lower) table shows the results in the NH (IH).}
	\label{m_CP-phases}
  \end{center}
\end{table}

%%%%%%%%%%%%%%%%%%%%%%%%%%%%%%%%%%%%%%%%%%%%%
\subsection{RG Evolution of the Mass Squared Differences} \label{MassSquared}
%%%%%%%%%%%%%%%%%%%%%%%%%%%%%%%%%%%%%%%%%%%%%

We show the RG evolution of the mass eigenvalues in this subsection and the mixing angles in the next subsection.
As mentioned above, the RG evolution of the masses depends on both $r$ and $\epsilon$,
 while those of the mixing angles depend only on $\epsilon$.
In the SM or the MSSM with small $\tan \beta$,
 all mixing angles almost stable because of the smallness of $\epsilon$.
Thus, we do not consider these cases.
From here, all figures correspond to the results in the MSSM with $\tan \beta = 30$ and SUSY threshold is taken at $\mu =$ 1\,TeV.

Figure \ref{deltam} shows the RG evolution of the mass squared differences
 ($\Delta m_{21}^2 \equiv m_2^2 -m_1^2$ and $\Delta m_{31}^2 \equiv |m_3^2 -m_1^2|$) with the decoupling effects.
The regions between each type of lines can be allowed by arbitrary combination of three CP-phases.
For example, the shaded regions are the allowed region for $m_{1 {\rm \,(or\,3)}}$ = 0.07\,eV.
This value of $m_{1 {\rm \,(or\,3)}}$ corresponds to the upper bound imposed
 by Planck 2013 results (Planck + WP + highL + BAO),
 which is given by $\sum_i m_i \leq 0.23$\,eV \cite{Planck2013}.
We can see that, when $m_{1 {\rm \,(or\,3)}}$ becomes large,
 $\Delta m_{21}^2$ can drastically vary in high energy scale compared with $\Delta m_{31}^2$.
The reason can be understood by the RGEs of the mass squared differences, which are written by
\begin{eqnarray}
	&& \frac{d}{dt} \Delta m_{21}^2 = C_1 \Delta m_{21}^2 + C_2 m_1^2\,, \\
	&& \frac{d}{dt} \Delta m_{31}^2 = C_3 \Delta m_{31}^2 \pm C_4 m_1^2 \quad (+:{\rm NH}, -:{\rm IH})\,, \label{m31}
\end{eqnarray}
where $C's$ represent the corresponding coefficients.
These RGEs show the feature that the evolution of $\Delta m_{21}^2$ is more sensitive to the value of $m_1$
 (equivalently the neutrino mass degeneracy) than that of $\Delta m_{31}^2$,
 because of $\Delta m_{21}^2\ll \Delta m_{31}^2$.

Now we note the CP-phase dependences of $\Delta m_{21}^2$ and $\Delta m_{31}^2$.
When $m_1 = 0$\,eV in the NH, both $\Delta m_{21}^2$ and $\Delta m_{31}^2$ are independent of $\phi_1$,
 while when $m_3 = 0$\,eV in the IH, they are independent of $|\phi_1-\phi_2|$.
The reason is because, in the light neutrino mass matrix,
 the mass eigenvalues are always appeared as $\left( m_1 e^{i \phi_1},\ m_2 e^{i \phi_2},\ m_3 \right)$
 (see Eq.\,(\ref{lM})).
When $m_{1 {\rm \,(or\,3)}} \neq 0$\,eV, the upper and lower parts of the allowed regions,
 except $\Delta m_{31}^2$ in the IH, are taken by $\delta =$ 0 and $\pi$, respectively.
For $\Delta m_{31}^2$ in the IH, they are taken by $\delta = \pi$ and 0, respectively.
The reason why this case is different from the others can be understood by Eq.\,(\ref{m31}).
In the right-hand side of this equation, the sign of term which is proportional to $m_1^2$ depends on the mass spectrum,
 since $\Delta m_{31}^2$ is defined as the absolute value, that is, $\Delta m_{31}^2 \equiv |m_3^2 - m_1^2|$.
Therefore, $\Delta m_{31}^2$ in the IH inversely behaves compared with that in the NH.
In particular, the upper and lower bounds are taken by some combinations of CP-phases as Table \ref{m_CP-phases}.

\begin{figure}[t]
  \begin{center}
      \begin{minipage}{0.43\hsize}
        \begin{center}
          \includegraphics[clip, width=\hsize]{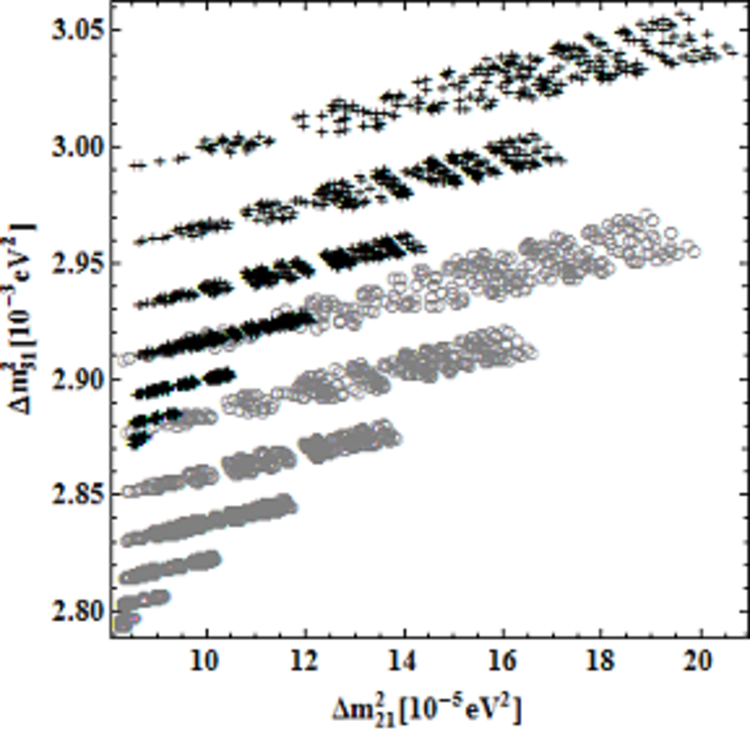}
        \end{center}
      \end{minipage}
\hspace{7mm}
      \begin{minipage}{0.43\hsize}
        \begin{center}
          \includegraphics[clip, width=\hsize]{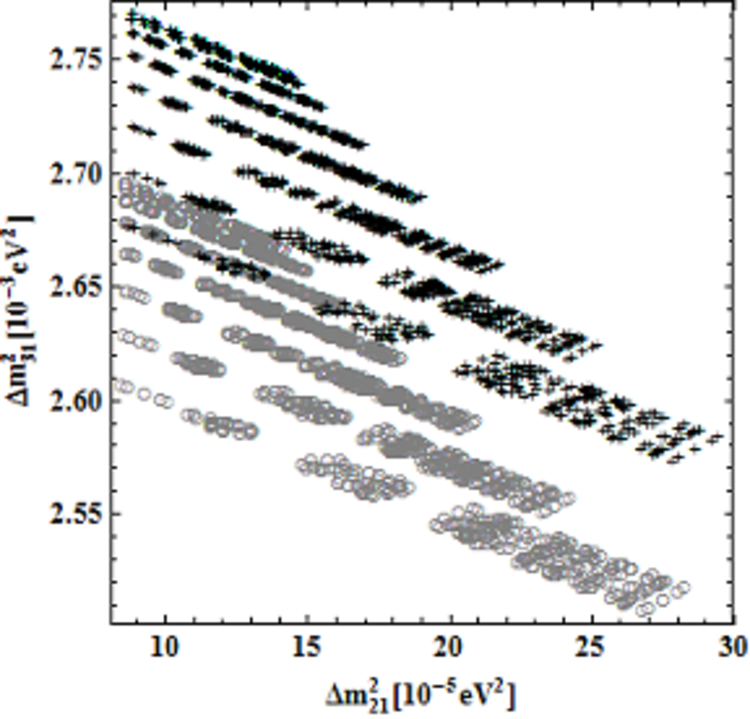}
        \end{center}
      \end{minipage}
    \caption{$\Delta m_{21}^2$ vs. $\Delta m_{31}^2$ at $\mu = 10^{14}$\,GeV.
				The left (right) figure shows the results in the NH (IH).
				"$\bigcirc$"(gray)  and "+"(black) represent the results including the decoupling effects and not, respectively.
				The clusters correspond to $m_{1 {\rm \,(or\,3)}}$ = 0\,eV, 0.01\,eV, 0.02\,eV, $\cdots$, and 0.07\,eV from the bottom (top) in the NH (IH).}
    \label{deltam_deltam}
  \end{center}
\end{figure}

Figure \ref{deltam_deltam} shows $\Delta m_{21}^2$ vs. $\Delta m_{31}^2$ at $\mu = 10^{14}$\,GeV.
As seen in Fig.\ref{deltam}, the allowed regions are large for large $m_{1 {\rm \,(or\,3)}}$.
The gradients of the figures reflect the sign of term which is proportional to $m_1^2$ in Eq.\,(\ref{m31}).
We can see that the differences between including the decoupling effects or not
 are about 3.5\% (4.0\%) for $\Delta m_{21}^2$, and 2.9\% (2.7\%) for $\Delta m_{31}^2$ in the NH (IH).
These magnitudes of the differences are nearly the same for any CP-phases.
When we construct the models in high energy scale,
 to reproduce the experimental values in low energy scale,
 we should input the parameters within the allowed regions shown in Fig.\ref{deltam}.
Figure \ref{deltam_deltam} shows the correct allowed parameters are about 3\% lower
 than the allowed parameters without the decoupling effects.

% \newpage
%%%%%%%%%%%%%%%%%%%%%%%%%%%%%%%%%%%%%%%%%%%%%
\subsection{RG Evolution of the Mixing Angles} \label{MixingAngles}
\begin{figure}[!ht]
  \begin{center}
      \begin{minipage}{0.32\hsize}
        \begin{center}
          \includegraphics[clip, width=\hsize]{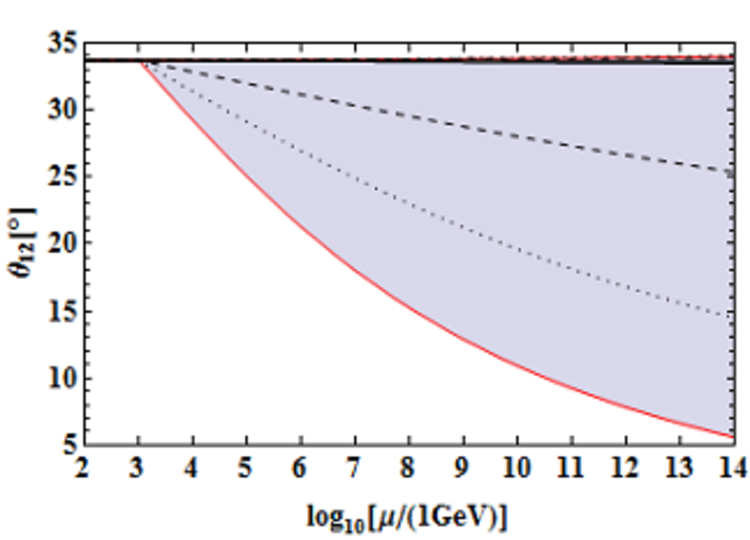}
        \end{center}
      \end{minipage}
\hspace{0.5mm}
      \begin{minipage}{0.32\hsize}
        \begin{center}
          \includegraphics[clip, width=\hsize]{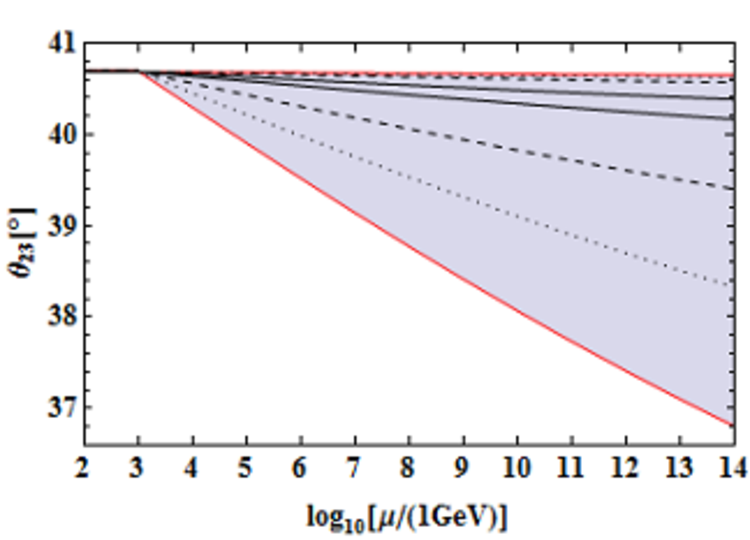}
        \end{center}
      \end{minipage}
\hspace{0.5mm}
      \begin{minipage}{0.32\hsize}
        \begin{center}
          \includegraphics[clip, width=\hsize]{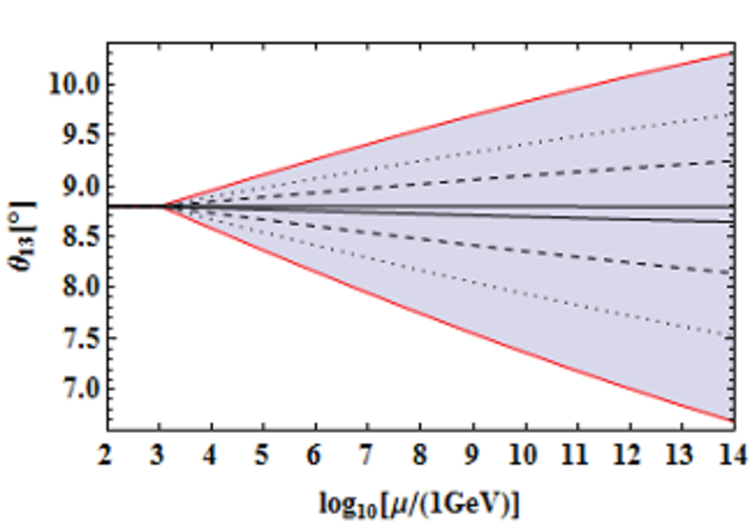}
        \end{center}
      \end{minipage}
  \end{center}
% \end{figure}
\vspace{-6mm}
% \begin{figure}[ht]
  \begin{center}
      \begin{minipage}{0.32\hsize}
        \begin{center}
          \includegraphics[clip, width=\hsize]{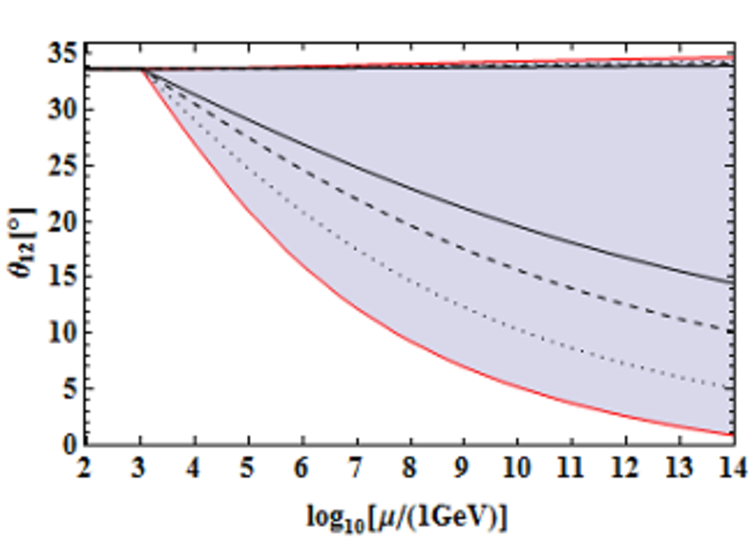}
        \end{center}
      \end{minipage}
\hspace{0.5mm}
      \begin{minipage}{0.32\hsize}
        \begin{center}
          \includegraphics[clip, width=\hsize]{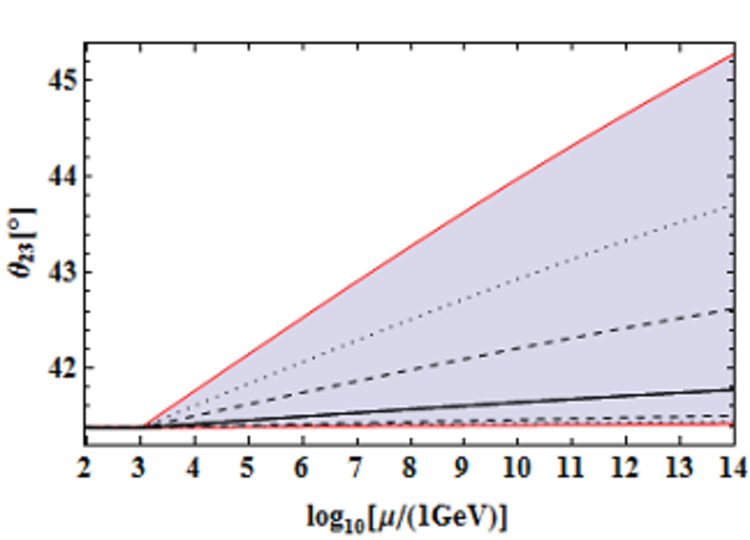}
        \end{center}
      \end{minipage}
\hspace{0.5mm}
      \begin{minipage}{0.32\hsize}
        \begin{center}
          \includegraphics[clip, width=\hsize]{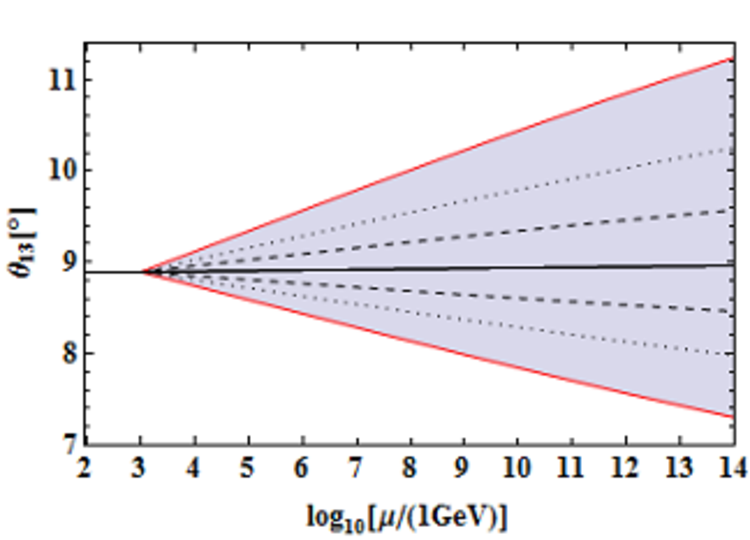}
        \end{center}
      \end{minipage}
    \caption{RG evolution of the mixing angles in the MSSM with the decoupling effects.
				The upper (lower) figures show the results in the NH (IH).
				The solid, dashed, dotted, and red-solid lines correspond to the upper and lower bound of allowed region
				 for $m_{1 {\rm \,(or\,3)}}$ = 0\,eV, 0.03\,eV, 0.05\,eV, and 0.07\,eV, respectively.
				The shaded regions can be taken according to CP-phases for $m_{1 {\rm \,(or\,3)}}$ = 0.07\,eV.}
    \label{mixing_angles}
  \end{center}
\end{figure}

%%%%%%%%%%%%%%%%%%%%%%%%%%%%%%%%%%%%%%%%%%%%%
Figure \ref{mixing_angles} shows the RG evolution of the mixing angles ($\theta_{12}$, $\theta_{23}$, and $\theta_{13}$) with the decoupling effects.
The settings of Fig.\ref{mixing_angles} are the same as in Fig.\ref{deltam}.
We can see that the allowed regions of all mixing angles are larger
 as $m_{1 {\rm \,(or\,3)}}$ is large, that is, the mass degeneracy is strong.
Particularly, the allowed region of $\theta_{12}$ is much larger than the others,
 since only $\theta_{12}$ strongly depends on $\Delta m_{21}^2$, %see hep-ph/0305273 (Table2)
 which can be unstable for large $m_{1 {\rm \,(or\,3)}}$.
On the other hand, $\theta_{23}$ and $\theta_{13}$ depend on rather $\Delta m_{31}^2$.
Note that the decoupling effects are negligible for the mixing angles,
 since they depend only on $\epsilon$ and the decoupling effects for $\epsilon$ are negligible
 as we have seen in Sec.\,\ref{MainResults}.

Finally, we comment on the CP-phase dependences of the mixing angles.
When $m_1 = 0$\,eV in the NH, all mixing angles are independent of $\phi_1$.
When $m_3 = 0$\,eV in the IH, $\theta_{12}$ is independent of $|\phi_1-\phi_2|$,
 and $\theta_{23}$ and $\theta_{13}$ are almost independent of all CP-phases.
The reasons can be almost understood by the same explanation as the cases of the mass squared differences,
 that is, in the light neutrino mass matrix,
 the mass eigenvalues are always appeared as $\left( m_1 e^{i \phi_1},\ m_2 e^{i \phi_2},\ m_3 \right)$.
In addition, when $m_3$ is small, $\theta_{23}$ are suppressed by $m_3$ and $\theta_{13}$ are stable \cite{Antusch:2003kp}.
When $m_{1 {\rm \,(or\,3)}} \neq 0$\,eV,
 the upper and lower parts of the allowed regions for $\theta_{12}$ are taken
 by $|\phi_1 - \phi_2| =$ $\pi$ and 0, respectively.
For $\theta_{23}$ in the NH (IH), the upper and lower (lower and upper) parts are taken
 by $(\phi_1, \phi_2)=$ ($\pi$, $\pi$) and (0, 0), respectively.
For $\theta_{13}$ in the NH (IH), the upper and lower (lower and upper) parts are taken 
 by $(|\delta - \phi_1|, |\delta - \phi_2|)=$ (0, $\pi$) and ($\pi$, 0), respectively.
Particularly, the upper and lower bounds are taken by some combinations of the CP-phases as Table \ref{th_CP-phases} and \ref{th12_CP-phases}.

\newpage

\begin{table}[!ht]
  \begin{center}
	\begin{tabular}{|c||c|c|c|} \hline
		NH & $\theta_{12}$ & $\theta_{23}$ & $\theta_{13}$ \\ \hline \hline
		Upper bound & depend on $m_1$ & (0, any, $\pi$) / (0, $\pi$, $\pi$) & ($\pi$, any, 0) / ($\pi$, $\pi$, 0) \\ \hline
		Lower bound & ($\pi$, any, 0) / ($\pi$, $\pi$, $\pi$) & (0, any, 0) / (0, 0, 0) & (0, any, 0) / ($\pi$, 0, $\pi$) \\ \hline
	\end{tabular}\\
\vspace{3mm}
	\begin{tabular}{|c||c|c|c|} \hline
		 IH & $\theta_{12}$ & $\theta_{23}$ & $\theta_{13}$ \\ \hline \hline
		Upper bound & depend on $m_3$ & \quad - \quad / ($\pi$, 0, 0) & \quad - \quad / ($\pi$, 0, $\pi$) \\ \hline
		Lower bound & $\delta = \pi, |\phi_1 - \phi_2| = 0$ / ($\pi$, 0, 0) & \quad - \quad / ($\pi$, $\pi$, $\pi$) & \quad - \quad / ($\pi$, $\pi$, 0) \\ \hline
	\end{tabular}
	\caption{Combinations of CP-phases which give the upper and lower bounds of $\theta_{12}$, $\theta_{23}$ and $\theta_{13}$.
	The values in the table are ($\delta, \phi_1, \phi_2$),
	 and the former and latter combinations correspond to $m_{1 {\rm \,(or\,3)}}$ = 0\,eV and nonzero $m_{1 {\rm \,(or\,3)}}$, respectively.
	The upper (lower) table shows the results in the NH (IH).
	"-" represents independence of CP-phases.}
	\label{th_CP-phases}
  \end{center}
% \end{table}
%
% \begin{table}[ht]
\vspace{-3mm}
  \begin{center}
	\begin{tabular}{|c||c|c|c|c|} \hline
		$m_1$ (or $m_3)$ & 0\,eV & 0.03\,eV & 0.05\,eV & 0.07\,eV \\ \hline \hline
		\multirow{4}{24mm}{Upper bound of $\theta_{12}$} & \multirow{2}{19mm}{(0, any, 0)} & \multirow{2}{15mm}{(0, $\pi$, 0)} & $(\frac{\pi}{2}, \frac{3\pi}{2}, \frac{\pi}{2})$ & $(\frac{\pi}{2}, \frac{\pi}{2}, \frac{3\pi}{2})$ \\
		& & & or $(\frac{3\pi}{2}, \frac{\pi}{2}, \frac{3\pi}{2})$ & or $(\frac{3\pi}{2}, \frac{3\pi}{2}, \frac{\pi}{2})$ \\
		\hhline{|~----|}
		& $\delta = \frac{\pi}{2}$ or $\frac{3\pi}{2}$, & $(\frac{\pi}{2}, \frac{\pi}{2}, \frac{3\pi}{2})$ & $(\frac{\pi}{2}, 0, \pi)$ & $(\frac{\pi}{2}, 0, \pi)$ \\
		& $|\phi_1 - \phi_2| = \pi$ & or $(\frac{3\pi}{2}, \frac{3\pi}{2}, \frac{\pi}{2})$ & or $(\frac{3\pi}{2}, 0, \pi)$ & or $(\frac{3\pi}{2}, 0, \pi)$ \\ \hline
	\end{tabular}
	\caption{Upper bound for $\theta_{12}$.
				The upper and lower combinations are corresponding to the NH and the IH, respectively.}
	\label{th12_CP-phases}
  \end{center}
\end{table}

\newpage
%%%%%%%%%%%%%%%%%%%%%%%%%%%%%%%%%%%%%%%%%%%%%%%%%%%%%%%%%%%%%%%%%%%%%%%%%%%
\section{Summary} \label{Summary}
%%%%%%%%%%%%%%%%%%%%%%%%%%%%%%%%%%%%%%%%%%%%%%%%%%%%%%%%%%%%%%%%%%%%%%%%%%%
We have investigated accurate renormalization group analyses in neutrino sector
 between $\nu$-oscillation and seesaw energy scales.
In other words, we have considered the decoupling effects of top quark and Higgs boson
 on the RGEs of the light neutrino mass matrix.
%  and shown the differences between including the effects or not.
Since the decoupling effects are given in the SM scale
 and independent of high energy physics,
 our method can basically apply to any models beyond the SM.
Therefore, it is useful to use our method when one constructs the models in high energy scale.
The relevant RGEs of the work are shown in Appendix.

In our analyses, we have used the effective dimension five operator, %which is called Weinberg operator.
 then the light neutrino mass matrix is approximately described only with two parameters,
 that is, $r$ and $\epsilon$.
$r$ is the overall factor of the mass matrix, and $\epsilon$ denotes the RG effects of charged lepton Yukawa couplings.
Using these parameters, the mass eigenvalues depend on both $r$ and $\epsilon$,
 while the mixing angles depend only on $\epsilon$.
We have shown the decoupling effects of top quark and Higgs boson for these parameters.
The effects of Higgs boson have been negligible,
 but those of top quark have been considerable because of the largeness of top quark Yukawa coupling.
For $r$, the differences between including the decoupling effects or not
 have been about 0.6\% in the SM and 1.4\% in the MSSM at $\mu = 10^{14}$\,GeV.
On the other hand, the differences for $\epsilon$ have been negligible in all energy scale,
 since $\epsilon$ depends on the integral of charged lepton Yukawa couplings,
 and then the decoupling effects are buried in the integrated value.
In the MSSM, the differences between including the decoupling effects or not
 are almost independent of the SUSY threshold and $\tan \beta$.
Besides, when $\tan \beta \simeq 1$, the RG evolution has been similar to the SM results.
These all results have been independent of the mass spectrum of the light neutrinos, the mass degeneracy and all CP-phases.
In other words, both $r$ and $\epsilon$ do not depend on the absolute neutrino mass scale and all CP-phases.

Next, we have shown the decoupling effects for the mass squared differences and the mixing angles.
Once we calculate $r$ and $\epsilon$,
 we can obtain the mass eigenvalues and the mixing angles by extracting from the light neutrino mass matrix.
We have derived the differences between including the decoupling effects or not
 are about 3.5\% (4.0\%) for $\Delta m_{21}^2$, and 2.9\% (2.7\%) for $\Delta m_{31}^2$ at $\mu = 10^{14}$\,GeV in the NH (IH).
These magnitudes of the differences have been nearly the same for any CP-phases.
Since the mixing angles depend only on $\epsilon$ and the differences for $\epsilon$ has been negligible,
 the differences for the mixing angles have been also negligible.

The RG analyses can be applied to some phenomenological problems,
 e.g. neutrinoless double beta decay or thermal leptogenesis, which were discussed in Ref.\,\cite{Antusch:2003kp}.
The amplitude of neutrinoless double beta decay is proportional to $(M_\nu)_{11}$, which is given by $r \times (M_\nu (M_Z))_{11}$.
In the leptogenesis, the heaviest mass eigenvalue and the absolute neutrino mass are the parameters
 used to calculate the baryon asymmetry.
Thus, accurate RG analyses are important to study these problems.
Note that our analyses correct the previous results, and the corrections would be not negligible.

% \newpage
%%%%%%%%%%%%%%%%%%%%%%%%%%%%%%%%%%%%%%%%%%%%%%%%%%%%%%%%%%%%%%%%%%%%%%%%%%%
\subsection*{\centering Acknowledgment} \label{Acknowledgement}
%%%%%%%%%%%%%%%%%%%%%%%%%%%%%%%%%%%%%%%%%%%%%%%%%%%%%%%%%%%%%%%%%%%%%%%%%%%
This work is partially supported by Scientific Grant by Ministry of Education, Culture, Sports, Science, and Technology, 
 Nos. 24540272, 20244028, 21244036, 23340070, and by the SUHARA Memorial Foundation.
The works of K.K, R.T., and Y.Y. are supported by Research Fellowships of the Japan Society
 for the Promotion of Science for Young Scientists (No. 24.801 (R.T.), 24E1077 (K.K), and 26E2428 (Y.Y.)).
The work is also supported by World Premier International Research Center Initiative (WPI Initiative), MEXT, Japan.

% \newpage
%%%%%%%%%%%%%%%%%%%%%%%%%%%%%%%%%%%%%%%%%%%%%%%%%%%%%%%%%%%%%%%%%%%%%%%%%%%
\section*{Appendix\quad Renormalization Group Equations} \label{app:RGEs}
%%%%%%%%%%%%%%%%%%%%%%%%%%%%%%%%%%%%%%%%%%%%%%%%%%%%%%%%%%%%%%%%%%%%%%%%%%%
\appendix
\setcounter{section}{1}
\setcounter{equation}{0}

In order to solve the RGEs of the coefficient of effective dimension five operator,
 the RGEs for all the parameters of the theory have to be solved simultaneously.
We summarize the RGEs for the extended SM and the extended MSSM.

%%%%%%%%%%%%%%%%%%%%%%%%%%%%%%%%%%%%%%%%%%%%%
\subsection{The RGEs of the Gauge Couplings} \label{app:RGEsgauge}
%%%%%%%%%%%%%%%%%%%%%%%%%%%%%%%%%%%%%%%%%%%%%
The RGEs of the gauge couplings are given by 
\begin{eqnarray}
	16 \pi^2 \, \beta_{g_A} \equiv 16 \pi^2 \, \frac{{\rm d} g_A}{{\rm d} t} = b_A \, g_A^3 \, ,
\end{eqnarray} 
 with
\begin{subequations}
\begin{eqnarray}
	b_1 &=& \frac{2}{5} \left[ \left(\frac{1}{6}\right)^2 6 N_Q + \left(\frac{2}{3}\right)^2 3 N_U + \left(\frac{1}{3}\right)^2 3 N_D + \left(\frac{1}{2}\right)^2 2 N_L + N_E \right]\nonumber \\ &&+ \frac{1}{5} \left(\frac{1}{2}\right)^2 2 N_H\, ,\\
	b_2 &=& - \frac{11}{3}\, 2 + \frac{1}{3} (3 N_Q + N_L) + \frac{1}{6} N_H\,,\\
	b_3 &=& - \frac{11}{3}\, 3 + \frac{1}{3} (2 N_Q + N_U + N_D)\,,
\end{eqnarray} 
\end{subequations}
 in the SM and $(\frac{33}{5}, 1, -3)$ in the MSSM, respectively.
We use ${\rm U}(1)_{\rm Y}$ gauge coupling with GUT charge normalization.
%, $g_1^{\rm GUT} = \sqrt{\frac{5}{3}}\, g_1$.
$N's$ represent the numbers of generations which are effective on the RGEs, and are given by Table \ref{number}.

\begin{table}[ht]
  \begin{center}
	\begin{tabular}{|c|c|c|c|c|c|c|} \hline
		& $N_Q$ & $N_U$ & $N_D$ & $N_L$ & $N_E$ & $N_H$  \\ \hline
		$\mu \geq m_t^{\rm pole}$ & 3 & 3 & 3 & 3 & 3 & 1\\ \hline
		$m_h \leq \mu < m_t^{\rm pole}$ & 2 & 2 & 3 & 3 & 3 & 1\\ \hline
		$M_Z \leq \mu < m_h$ & 2 & 2 & 3 & 3 & 3 & 0\\ \hline
	\end{tabular}
\caption{Numbers of generations which are effective on the RGEs.}
\label{number}
\end{center}
\end{table}

\newpage
%%%%%%%%%%%%%%%%%%%%%%%%%%%%%%%%%%%%%%%%%%%%%
\subsection{The RGEs in the SM} \label{app:RGEsSM}
%%%%%%%%%%%%%%%%%%%%%%%%%%%%%%%%%%%%%%%%%%%%%
In the extended SM, we can consider the effective dimension five operator
 (the coefficient is denoted by $\kappa$) in low energy scale.
The RGEs without the decoupling effects are given by the following $\beta$-functions \cite{Lindner:2002, Lindner:1987a}: 
\begin{subequations}
\begin{eqnarray}
	16\pi^2 \beta_\kappa &=& - \frac{3}{2} (Y_E^\dagger Y_E)^T \, \kappa - \frac{3}{2}\,\kappa \, (Y_E^\dagger Y_E) 
		+ 2 \, {\rm Tr} \left[3 Y_U^\dagger Y_U + 3 Y_D^\dagger Y_D + Y_E^\dagger Y_E \right] \, \kappa \nonumber \\
		&& - 3 g_2^2\, \kappa + \lambda \kappa \, ,\label{kappa_SM}
\\
	16 \pi^2 \beta_{Y_U} &=& Y_U \left\{ \frac{3}{2} Y_U^\dagger Y_U - \frac{3}{2}\, Y_D^\dagger Y_D
		+ {\rm Tr} \left[3 Y_U^\dagger Y_U + 3 Y_D^\dagger Y_D + Y_E^\dagger Y_E \right] \right. \nonumber \\
		&& \quad \; \left. - \frac{17}{20} g_1^2 - \frac{9}{4} g_2^2 - 8 \,g_3^2 \right\} ,\label{yu_SM}
\\
	16 \pi^2 \beta_{Y_D} &=& Y_D \left\{ \frac{3}{2} Y_D^\dagger Y_D - \frac{3}{2}\, Y_U^\dagger Y_U
		+ {\rm Tr} \left[3 Y_U^\dagger Y_U + 3 Y_D^\dagger Y_D + Y_E^\dagger Y_E \right] \right. \nonumber \\
		&& \quad \; \left. - \frac{1}{4} g_1^2 - \frac{9}{4} g_2^2 - 8 \,g_3^2 \right\} ,
\\
	16 \pi^2 \beta_{Y_E} &=& Y_E \left\{ \frac{3}{2} Y_E^\dagger Y_E
		+ {\rm Tr} \left[3 Y_U^\dagger Y_U + 3 Y_D^\dagger Y_D + Y_E^\dagger Y_E \right] 
		- \frac{9}{4} g_1^2 - \frac{9}{4} g_2^2 \right\} ,
\\
	16 \pi^2 \beta_{\lambda} &=& 6 \lambda^2 - \left( \frac{9}{5} g_1^2 + 9 g_2^2 \right) \lambda
							+ \frac{9}{2} \left( \frac{3}{25} g_1^4 + \frac{2}{5} g_1^2 g_2^2 + g_2^4 \right) \nonumber  \\
							&& + 4 \, {\rm Tr} \left[ 3 Y_U^\dagger Y_U + 3 Y_D^\dagger Y_D + Y_E^\dagger Y_E \right] \lambda \nonumber \\
							&& - 8 \, {\rm Tr} \left[ 3 Y_U^\dagger Y_U Y_U^\dagger Y_U + 3 Y_D^\dagger Y_D Y_D^\dagger Y_D + Y_E^\dagger Y_E Y_E^\dagger Y_E \right] \, .
\end{eqnarray}
\end{subequations}
Here, the Higgs potential is given by $V(\phi) = -\frac{m_h^2}{2} |\phi|^2 + \frac{\lambda}{4} |\phi|^4$.
Then, $\lambda = \frac{m_h^2}{v^2}$, where $m_h$ is the mass of Higgs boson, and we take $m_h=126$\,GeV at $\mu = M_Z$ and $v=174$\,GeV.
We use these RGEs for $m_t^{\rm pole} \leq \mu$ ($<$ SUSY threshold). 

For $m_h \leq \mu < m_t^{\rm pole}$, top quark is decoupled, and $\beta$-functions are given as follows: 
\begin{subequations}
\begin{eqnarray}
	16\pi^2 \beta_\kappa &=& - \frac{3}{2} (Y_E^\dagger Y_E)^T \, \kappa - \frac{3}{2}\,\kappa \, (Y_E^\dagger Y_E) \nonumber \\
		&& + 2 \, \left( {\rm Tr} \left[3 Y_U^\dagger Y_U + 3 Y_D^\dagger Y_D + Y_E^\dagger Y_E \right] - 3 y_t^2 \right) \, \kappa %\nonumber \\
		 - 3 g_2^2\, \kappa + \lambda \kappa \, ,\label{kappa_mt}
\\
% 	16 \pi^2 \beta_{y_t} &=& y_t \left\{ - \frac{3}{2}\, y_b^2
% 		+ \left( {\rm Tr} \left[3 Y_U^\dagger Y_U + 3 Y_D^\dagger Y_D + Y_E^\dagger Y_E \right] - 3 y_t^2 \right) 
% 		- \frac{9}{20} g_1^2 - \frac{9}{4} g_2^2 \right\} , \nonumber \\
% \\
	16 \pi^2 \beta_{Y_U \in \{y_u,y_c\}} &=& Y_U \left\{ \frac{3}{2} Y_U^\dagger Y_U - \frac{3}{2}\, Y_D^\dagger Y_D
		+ \left( {\rm Tr} \left[3 Y_U^\dagger Y_U + 3 Y_D^\dagger Y_D + Y_E^\dagger Y_E \right] - 3 y_t^2 \right)  \right. \nonumber \\
		&& \quad \; \left. - \frac{17}{20} g_1^2 - \frac{9}{4} g_2^2 - 8 \,g_3^2 \right\} ,
\\
	16 \pi^2 \beta_{y_b} &=& y_b \left\{ \frac{3}{2} y_b^2
		+ \left( {\rm Tr} \left[3 Y_U^\dagger Y_U + 3 Y_D^\dagger Y_D + Y_E^\dagger Y_E \right] - 3 y_t^2 \right)  \right. \nonumber \\
		&& \quad \; \left. - \frac{1}{4} g_1^2 - \frac{9}{4} g_2^2 - 8 \,g_3^2 \right\} ,
\\
	16 \pi^2 \beta_{Y_D \in \{y_d,y_s\}} &=& Y_D \left\{ \frac{3}{2} Y_D^\dagger Y_D - \frac{3}{2}\, Y_U^\dagger Y_U
		+ \left( {\rm Tr} \left[3 Y_U^\dagger Y_U + 3 Y_D^\dagger Y_D + Y_E^\dagger Y_E \right] - 3 y_t^2 \right)  \right. \nonumber \\
		&& \quad \; \left. - \frac{1}{4} g_1^2 - \frac{9}{4} g_2^2 - 8 \,g_3^2 \right\} ,
\\
	16 \pi^2 \beta_{Y_E} &=& Y_E \left\{ \frac{3}{2} Y_E^\dagger Y_E
		+ \left( {\rm Tr} \left[3 Y_U^\dagger Y_U + 3 Y_D^\dagger Y_D + Y_E^\dagger Y_E \right] - 3 y_t^2 \right) 
		- \frac{9}{4} g_1^2 - \frac{9}{4} g_2^2 \right\} , \nonumber \\
\\
	16 \pi^2 \beta_{\lambda} &=& 6 \lambda^2 - \left( \frac{9}{5} g_1^2 + 9 g_2^2 \right) \lambda
							+ \frac{9}{2} \left( \frac{3}{25} g_1^4 + \frac{2}{5} g_1^2 g_2^2 + g_2^4 \right) \nonumber  \\
							&& + 4 \, \left( {\rm Tr} \left[ 3 Y_U^\dagger Y_U + 3 Y_D^\dagger Y_D + Y_E^\dagger Y_E \right] - 3 y_t^2 \right) \lambda \nonumber \\
							&& - 8 \, \left( {\rm Tr} \left[ 3 Y_U^\dagger Y_U Y_U^\dagger Y_U + 3 Y_D^\dagger Y_D Y_D^\dagger Y_D + Y_E^\dagger Y_E Y_E^\dagger Y_E \right] - 3 y_t^4 \right)\, .\label{lambda_mt}
\end{eqnarray}
\end{subequations}
Since $\beta$-function of top quark Yukawa coupling is not necessary for $\mu < m_t^{\rm pole}$, we omit it.
The decoupling effects of top quark are shown as $-3 y_t^2$ or $-3 y_t^4$,
 which cancel top quark Yukawa coupling in Tr[$Y_U^\dagger Y_U$] or Tr[$Y_U^\dagger Y_U Y_U^\dagger Y_U$].
Therefore, $\beta$-functions do not include top quark Yukawa coupling for $\mu < m_t^{\rm pole}$.

For $M_{\rm Z} \leq \mu < m_h$, Higgs boson is also decoupled, and $\beta$-functions are given as follows: 
\begin{subequations}
\begin{eqnarray}
	16\pi^2 \beta_\kappa &=& - 3 g_2^2\, \kappa \, ,\label{kappa_mh}
\\
% 	16 \pi^2 \beta_{y_t} &=& 0 \, ,
% \\
	16 \pi^2 \beta_{Y_U \in \{y_u,y_c\}} &=& Y_U \left( - \frac{2}{3} g_1^2 - 8 \,g_3^2 \right) ,\label{yu_mh}
\\
	16 \pi^2 \beta_{Y_D} &=& Y_D \left( \frac{1}{5} g_1^2 - 8 \,g_3^2 \right) ,
\\
	16 \pi^2 \beta_{Y_E} &=& Y_E \left( - \frac{9}{5} g_1^2 \right) ,\label{ye_mh}
\\
	16 \pi^2 \beta_{\lambda} &=& - 8 \, \left( {\rm Tr} \left[ 3 Y_U^\dagger Y_U Y_U^\dagger Y_U + 3 Y_D^\dagger Y_D Y_D^\dagger Y_D + Y_E^\dagger Y_E Y_E^\dagger Y_E \right] - 3 y_t^4 \right)\, .\label{lambda_mh}
\end{eqnarray}
\end{subequations}
In this energy region, Higgs boson also does not appear as the internal line in Feynman diagrams.
Then, Eq.(\ref{kappa_mh}) has only one term which is proportional to SU(2) gauge coupling,
 and Eq.(\ref{lambda_mh}) corresponds to fermion box diagrams.
Using Landau gauge, Eqs.(\ref{yu_mh})-(\ref{ye_mh}) are calculated by the right diagram in Fig.\ref{diagram}, which has U(1) gauge boson.

% Then, $\beta$-function of $\kappa$ has only one term which is proportional to SU(2) gauge coupling,
%  and $\beta$-function of $\lambda$ has only contributions of fermion box diagrams,
%  which appear as fourth power of Yukawa couplings.
% For $\beta$-functions of fermion Yukawa couplings,
%  the terms of gauge couplings remain.
% In order to calculate contributions of electroweak gauge bosons,
%  we use Landau gauge, in which only two diagrams shown in Fig.\ref{diagram} have nonzero contributions.
% Particularly, for $M_Z \leq \mu < m_h$
%  we have to calculate only the right figure,
%  which has U(1) gauge boson.

% \newpage
%%%%%%%%%%%%%%%%%%%%%%%%%%%%%%%%%%%%%%%%%%%%%
\subsection{The RGEs in the MSSM}\label{app:RGEsMSSM}
%%%%%%%%%%%%%%%%%%%%%%%%%%%%%%%%%%%%%%%%%%%%%
In the MSSM, we can consider the effective dimension five operator
 (the coefficient is denoted by $\kappa$) in low energy scale.
The decoupling effects of the massive SM particle do not affect the RGEs in the MSSM scale.
The RGEs are given by the following $\beta$-functions within the 1-loop level \cite{Lindner:2002, Lindner:1987b}:
\begin{subequations}
\begin{eqnarray}
	16\pi^2 \beta_\kappa &=& (Y_E^\dagger Y_E)^T \, \kappa + \kappa \, (Y_E^\dagger Y_E)
% 		+ (Y^\dagger_\nu Y_\nu)^T \, \kappa + \,\kappa \, (Y^\dagger_\nu Y_\nu) \nonumber \\
		+ 2 \, {\rm Tr} \left[3 Y_U^\dagger Y_U \right] \kappa 
		- \frac{6}{5} g_1^2\, \kappa - 6 g_2^2\, \kappa \, ,\label{kappa_SUSY}
\\
	16 \pi^2 \beta_{Y_U} &=& Y_U \left\{ 3 Y_U^\dagger Y_U + Y_D^\dagger Y_D
		+ {\rm Tr} \left[3 Y_U^\dagger Y_U \right] - \frac{13}{15} g_1^2 - 3 g_2^2 - \frac{16}{3} g_3^2 \right\} ,
\\
	16 \pi^2 \beta_{Y_D} &=& Y_D \left\{ 3 Y_D^\dagger Y_D + Y_U^\dagger Y_U
		+ {\rm Tr} \left[3 Y_D^\dagger Y_D + Y_E^\dagger Y_E \right] - \frac{7}{15} g_1^2 - 3 g_2^2 - \frac{16}{3} \,g_3^2 \right\} ,
% \\
%    16 \pi^2 \beta_{Y_\nu} &=& Y_\nu \left\{ 3 Y^\dagger_\nu Y_\nu + Y_E^\dagger Y_E
% 		+ {\rm Tr} \left[3 Y_U^\dagger Y_U \right] - \frac{3}{5} g_1^2 - 3 g_2^2 \right\} ,
\\
	16 \pi^2 \beta_{Y_E} &=& Y_E \left\{ 3 Y_E^\dagger Y_E %+ Y_\nu^\dagger Y_\nu
		+ {\rm Tr} \left[3 Y_D^\dagger Y_D + Y_E^\dagger Y_E \right] - \frac{9}{5} g_1^2 - 3 g_2^2 \right\} \, .
\end{eqnarray}
\end{subequations}

%%%%%%%%%%%%%%%%%%%%%%%%%%%%%%%%%%%%%%%%%%%%%%%%%%%%%%%%%%%%%%%%%%%%%%%%%%%

\end{document}